\documentclass[a4paper,11pt]{article}
\pdfoutput=1 
\usepackage{jheppub} 
\usepackage{lineno}

\usepackage{amsmath}
\usepackage{amssymb}
\usepackage{graphicx}
\usepackage{subcaption}
\usepackage{bbold}
\usepackage{orcidlink}
\usepackage{bbding}
\usepackage{slashed}
\usepackage{float}
\usepackage{verbatim}
\usepackage{hyperref}
\usepackage[normalem]{ulem}

\usepackage{natbib}
\setcitestyle{square,numbers,sort&compress,comma}

\title{\boldmath Searching for type I seesaw mechanism in a two Heavy Neutral Leptons scenario at FCC-ee}

\author[a]{S. Ajmal\orcidlink{0000-0002-2726-2858},}
\author[b]{P. Azzi\orcidlink{0000-0002-3129-828X},}
\author[c,1]{S. Giappichini\note{Corresponding author.}\orcidlink{0009-0002-7694-641X},}
\author[c]{M. Klute\orcidlink{0000-0002-0869-5631},}
\author[a]{O. Panella\orcidlink{0000-0003-4262-894X},}
\author[c]{M. Presilla\orcidlink{0000-0003-2808-7315}}
\author[c]{and X. Zuo\orcidlink{0000-0002-0029-493X}}

\affiliation[a]{INFN Sezione di Perugia, Via A. Pascoli, I-06123, Perugia, Italy}
\affiliation[b]{INFN Sezione di Padova, Via Marzolo 8, I-35100, Padova, Italy}
\affiliation[c]{Institute for Experimental Particle Physics (ETP), Karlsruhe Institute of Technology (KIT), Wolfgang-Gaede-Straße 1, D-76131, Karlsruhe, Germany}

\abstract{
This paper investigates the search for heavy neutral leptons (HNL) in the type I seesaw mechanism at the Future Circular Collider in its $e^+e^-$ stage (FCC-ee), considering a luminosity of 125~ab$^{-1}$ collected at $\sqrt{s}=$91.2 GeV. The study examines two generations of heavy neutral leptons produced in association with Standard Model (SM) neutrinos and decaying to a purely leptonic final state.
This theoretical framework can explain neutrino oscillations and other open questions of the SM, providing a broader perspective on the relevance of this experimental search. The analysis is performed using a fast simulation of the IDEA detector concept to study potential HNL interactions at the FCC-ee. 
The sensitivity contours are obtained from a selection of kinematic variables aimed at improving the signal-to-background ratio for the prompt production case. In the case of long-lived HNLs, the background can be almost fully eliminated by exploiting their displaced decay vertices. The study shows that the FCC-ee has a significant sensitivity to observing these objects in a region of the phase space not accessible by other experiments. 
}

\begin{document}

\maketitle

\flushbottom

\section{Introduction}

The Standard Model (SM) provides the underlying framework for the current understanding of fundamental interactions, with the exception of the gravitational interaction. The remarkable agreement between the SM and the results of experiments at colliders makes it a powerful tool for predicting many high-energy processes. However, the SM does not explain several observed phenomena. These include neutrino oscillations, the observed baryon asymmetry of the Universe, the nature of dark matter, and the accelerated expansion of the Universe. Further insight into the origin of neutrino masses would facilitate a deeper understanding of particle masses and provide potential solutions to the remaining issues within the SM.
The SM assumes the existence of massless neutrinos based on the observation of negative helicity~\cite{helicity}. In contrast, solar and atmospheric neutrino experiments have demonstrated that neutrinos oscillate with two distinct squared-mass differences, as confirmed by the independent measurements conducted by the KamLAND and K2K experiments~\cite{kamland,k2k}. Consequently, the observed hierarchy can only be accommodated in two types of mixing schemes, normal and inverted, where the masses of the neutrinos are sorted by taking the lightest to be either the first or the third. More recently, KATRIN has found an upper limit of $m_\nu<0.45$ eV at 90\% confidence level~\cite{KATRIN}.
While this has resulted in the emergence of numerous beyond the SM scenarios, only a few of them are often selected for collider studies~\cite{CMS:2024bni}. Among these, the type I seesaw mechanism~\cite{typei,typei1,typei2} is of great interest and forms the basis of this study. The introduction of sterile heavy neutrinos (also referred to as heavy neutral leptons, HNLs) allows the generation of SM neutrino masses compatible with the two observed hierarchies and could account for the generation of the baryon asymmetry of the Universe via leptogenesis~\cite{baryogenesis}, proposing a suitable dark matter candidate~\cite{coldmatter, darkmatter1} when the number of HNLs is at least two. Moreover, by appropriately choosing the parameters, the HNLs can have electroweak-scale masses, which means they could be produced and indirectly observed by their decay products by current and future experiments~\cite{production,futurehnl,collider}. 

The proposed simulation environment is based on the future electron-positron collider, the FCC-ee (Future Circular Collider)~\cite{fccphysics,fcclepton}, which is currently under consideration for construction at CERN in the coming years. The large integrated luminosity of the FCC-ee runs, along with the wide range of explored energies, will facilitate searches for any small deviations from the SM predictions, as well as the observation of rare processes in a clean and efficient environment. 
This work explores events with two charged leptons and missing energy produced in electron-positron collisions at a center of mass energy of $\sqrt{s}=91.2$ GeV and corresponding luminosity of $\mathcal{L}=125$ ab$^{-1}$. The analysis of the simulated data exploits the performance of the Innovative Detector for Electron-positron Accelerators (IDEA)~\cite{fcclepton}, developed for FCC-ee. Its main components are a short-drift wire chamber and a dual-readout calorimeter, interleaved by a thin, low-mass superconducting solenoid coil. Its design makes it ideal for identifying the decay vertices of prompt particles as well as long-lived ones. 

In the context of FCC-ee physics studies with realistic detector simulations, only models considering a single HNL coupling to one flavor at a time have been explored~\cite{valle-polesello,snowmass, rygaard}. These studies showcase the unprecedented sensitivity that this future lepton collider will have for these scenarios. However, since the immediate consequence of detecting two distinct scales in neutrino oscillations is that the number of right-handed neutrinos must be at least two, the one-flavor scenario is unlikely to reflect the dynamics observed in nature. 
Consequently, one distinctive aspect of this study is the generalization of the underlying model's assumptions considering a more realistic scenario where two generations of quasi-degenerate HNLs couple to all lepton flavors.

\begin{figure}[h]
    \begin{subfigure}[h]{0.48\textwidth}
        \centering
        \label{fig:z}
        \includegraphics[width=\textwidth]{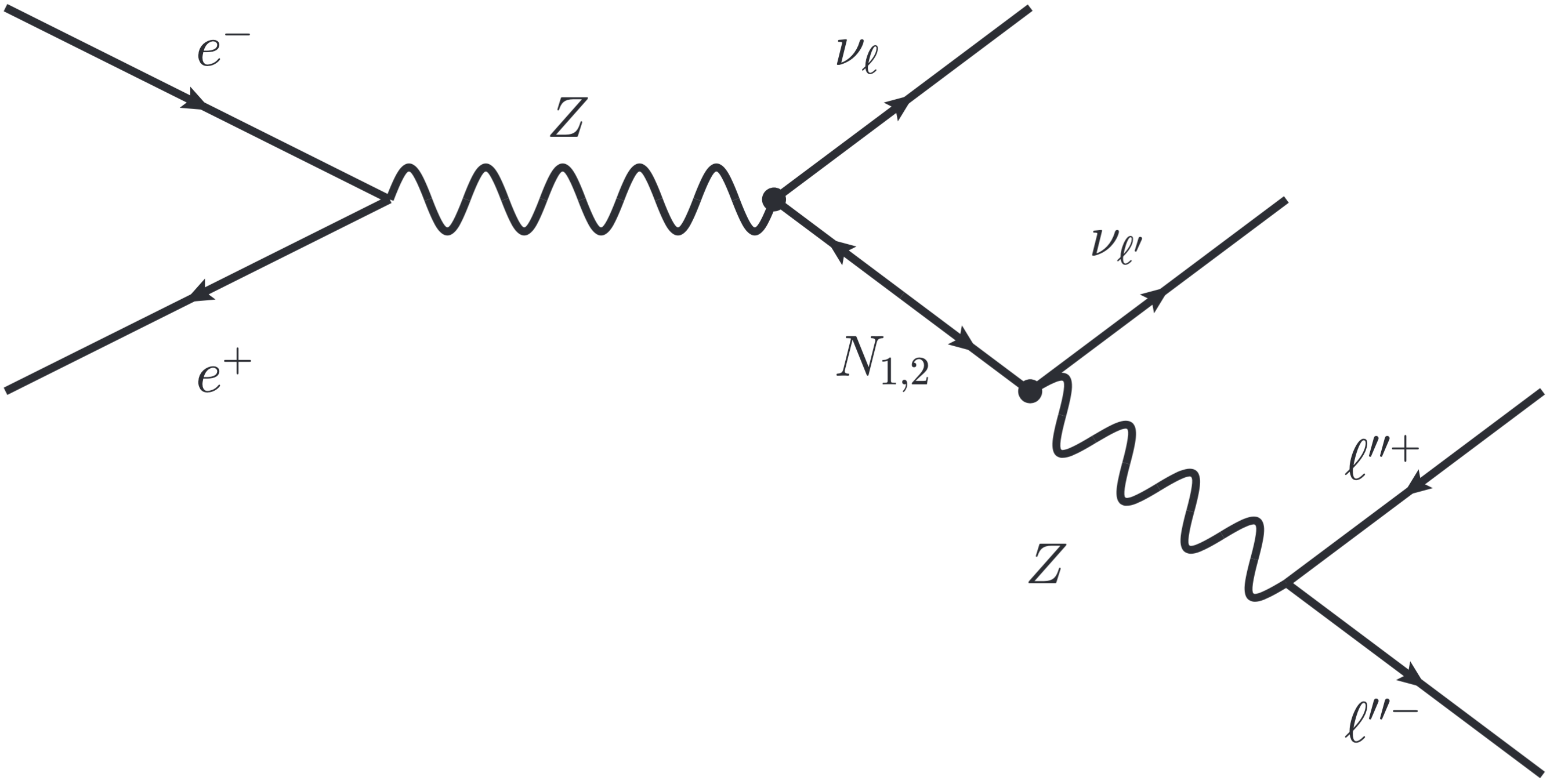}
    \end{subfigure}
    \hfill
    \begin{subfigure}[h]{0.48\textwidth}
        \centering
        \label{fig:w}
        \includegraphics[width=\textwidth]{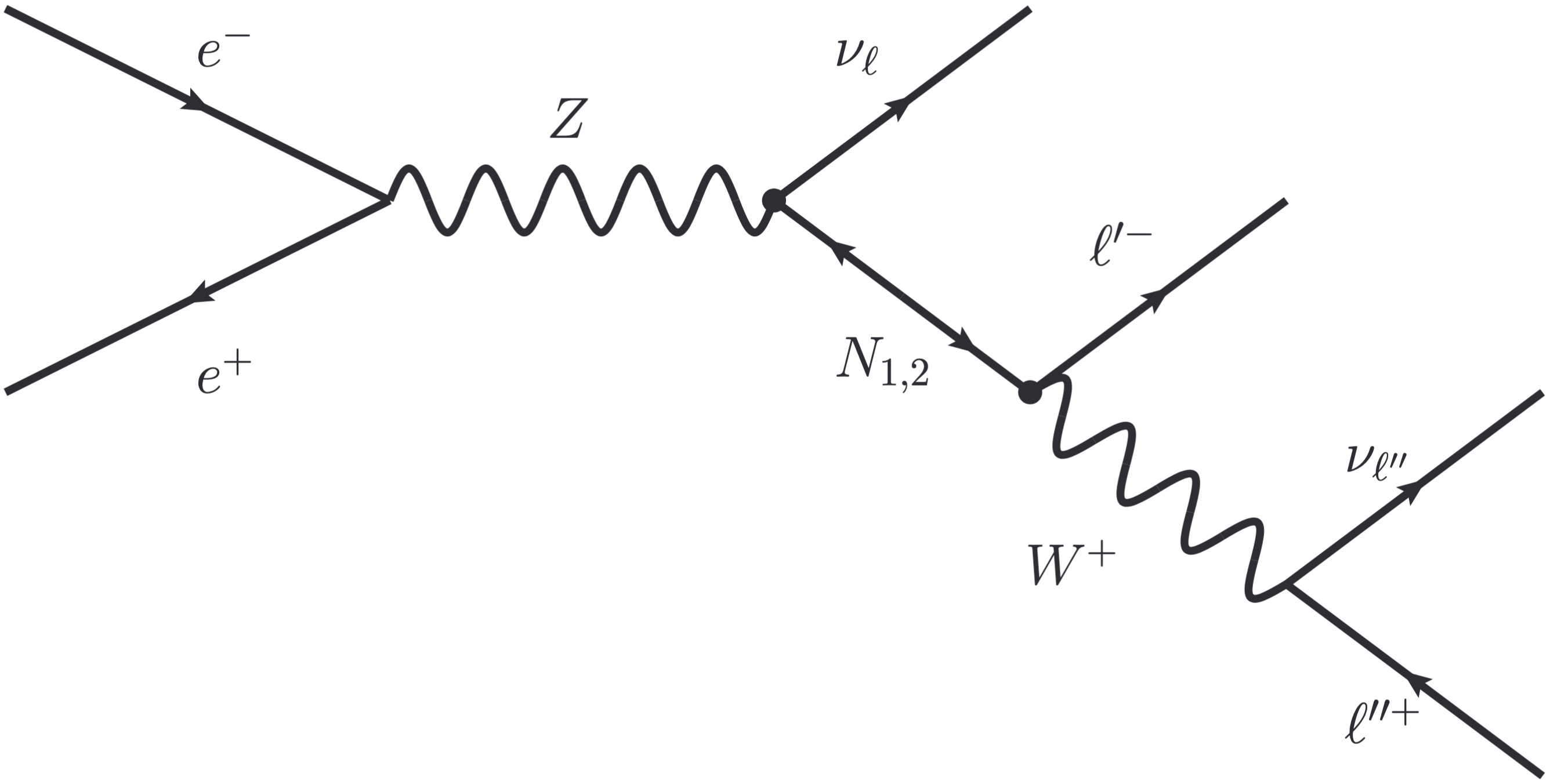}
    \end{subfigure}
    \caption{Feynman diagrams for the production of the two HNLs ($N_{1,2}$) associated with a light neutrino in a $e^+e^-$ collisions. On the left diagram, the HNL decay is mediated by a Z boson, and on the right by a W. The final state is in both cases composed of two SM neutrinos and two leptons. Two examples of lepton flavor and number violating diagrams are shown here.}
    \label{fig:diagrammi}
\end{figure}

Specifically, this work examines Majorana HNLs produced in association with a light neutrino from an s-channel Z boson from an electron-positron collision. The decay of the HNL is mediated by either a Z or W boson. Only their respective leptonic decays are considered in this study, restricting the search to the leptonic tau decays as well. The final state is then characterized by either two electrons or muons, or one electron and one muon, plus two neutrinos, as illustrated in Fig. \ref{fig:diagrammi}. 

The following is a description of the organization of the paper. In Section \ref{sec:MCsig}, the theory of the type I seesaw mechanism is introduced alongside a comment on the choices of the HNLs model adopted and the simulation setup used for the analysis.
Subsequently, in Section \ref{sec:bkg}, the modeling of background sources is discussed. The details of the analysis performed are given in Section \ref{sec:analysis} while the expected significance for both a shape-based inclusive analysis and background-free long-lived HNLs scenario is highlighted in Section \ref{sec:results}. A summary of the findings of this study concludes the paper in Section~\ref{sec:summary}.

\section{Theoretical model and simulation}
\label{sec:MCsig}

Adopting a realistic seesaw model for neutrino oscillations introduces several changes compared to the single-HNL model, including the possibility of accessing final states with different lepton flavors and additional free parameters. This analysis follows a theoretical scenario highlighted in Ref.~\cite{atlasrecast}, which investigates the case of two HNLs decaying into electrons and muons during Run 2 of the LHC with proton-proton collisions.
In the type I seesaw model, the new parameters generally describing the interactions between SM particles and a number $n$ of HNLs are one mass value $ M_i $ for each right-handed neutrino and a complex mixing matrix $ U_{\ell i} $ of dimension $ 3 \times n $ coupling HNLs to SM leptons~\cite{pascoli_lagrangian}. 
Supposing at least two heavy neutrinos, it is possible to consider a global $U(1)_{B-L}$ symmetry in which the light neutrinos become massless. If the symmetry is exact, the HNLs have to be organized in pairs with masses $M_i=M_j$ and mixing angles $U_{\ell i}=\text{i}U_{\ell j}$~\cite{Kersten_2007,Shaposhnikov_2007,pascoli}, forming a single Dirac particle.
Considering even a small mass splitting, $\Delta M_{ij}=|M_i-M_j|$ induced by symmetry breaking can give rise to lepton number violating (LNV) oscillations, making LNV processes unsuppressed when the HNL decay length exceeds the oscillation length. Since $\Delta M_{ij} \ll \overline{M}=(M_i+M_j)/2$ for the approximate symmetry to protect the masses, the two HNLs are Majorana-like particles, exhibiting both lepton number violating and conserving processes with the same integrated rate. This is referred to as the pseudo-Dirac limit.
Studying the type I see-saw mechanism in these conditions restricts the number of free parameters and enhances the cross-section of the process by allowing sizable mixing between active and sterile neutrinos. This option is also interesting since low-scale leptogenesis and dark matter production in the early Universe require a mass degeneracy between two heavy neutrinos~\cite{darkmatter_old,leptogenesis,Klaric_2021,Hern_ndez_2022,lgconstraints}. 
Additional constraints on HNL mixings can be imposed considering the unitarity of the mixing matrix by defining the ratios between mixing angles, \( U_\ell^2 = \sum_{i=1,..,n} U^2_{\ell i} \), and the total mixing angle, \( U^2 = \sum_{\ell=e,\mu,\tau} U^2_{\ell} \):
$$
\frac{U^2_{e}}{U^2} + \frac{U^2_\mu}{U^2} + \frac{U^2_\tau}{U^2} = 1.
$$
Neutrino oscillation data analyzed by the NuFIT collaboration~\cite{nufit} constrain the allowed ratios for HNL mixings consistent with a normal or inverted neutrino hierarchy. Ref.~\cite{bounds} conducted a global analysis of HNL parameters considering flavor and electroweak observables, validating the same regions. Including the requirement of generating sufficient leptogenesis, these regions can be further constrained as shown in Ref.~\cite{Klaric_2021, Hern_ndez_2022,lgconstraints}. FCC-ee will be capable of probing the HNL parameter space that meets all these conditions, as demonstrated in Ref.~\cite{leptogenesis,leptogenesis1}, making it well-suited for this search.

\begin{figure}[h]
    \centering
    \includegraphics[width=0.6\textwidth]{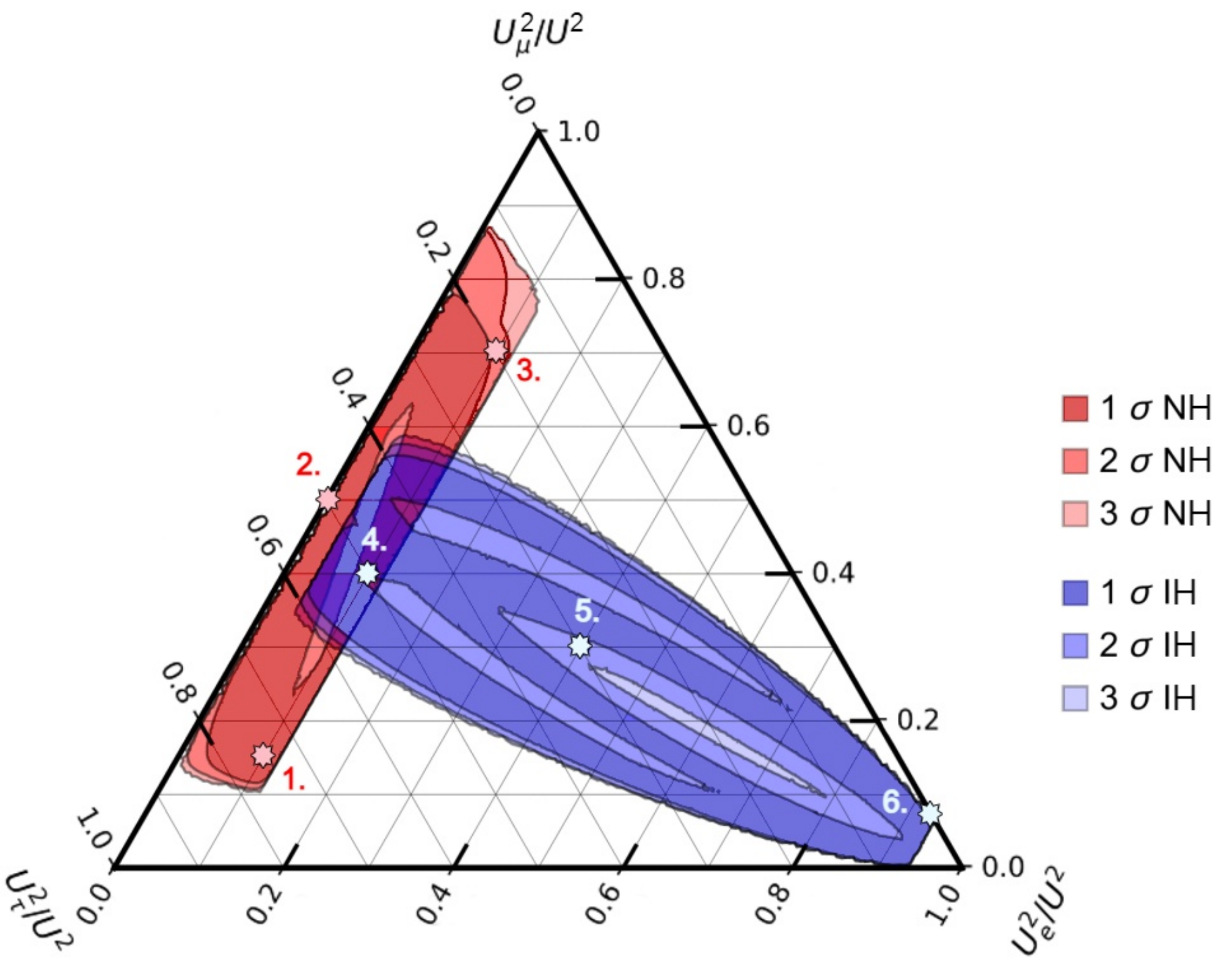}
    \caption{Illustration of ratios $U^2_{e}/U^2:U^2_\mu/U^2:U^2_\tau/U^2$ considered. The light pink stars represent HNLs consistent with the normal hierarchy, the light blue ones are for the inverted hierarchy. The range of allowed ratios is taken from the global fit on oscillation data by the NuFit collaboration~\cite{nufit, nufitweb}. The points chosen are also consistent with leptogenesis production~\cite{leptogenesis1}. Image adapted from Ref.~\cite{futurehnl}.}
    \label{fig:points}
\end{figure}

In this work, this scenario is addressed with a simulation of the HNL mediated process in the MC event generator for high-energy physics MadGraph5\_aMC v. 3.5.3~\cite{madgraph} integrated with the model \textit{HeavyN} at leading order~\cite{heavyn1, heavyn2} to introduce the HNLs interactions in the SM. Given the scope of this paper and the parameter space studied, this model, which does not simulate HNL oscillations, can describe the production and decay of two HNLs. Initial state radiation has not been taken into consideration at this stage. Subsequently, Pythia8 v. 8.306~\cite{pythia} has been used to hadronize the initial and final states and handle tau decays and Delphes v. 3.5.1pre05~\cite{delphes} to simulate the response of the IDEA detector. 
The value of the mass separation considered in this study is $\Delta M=|M_1-M_2|=1\times10^{-5}$ GeV to account for the constraints given by the cosmological arguments~\cite{Klaric_2021} while also enhancing the cross-section. To simultaneously create the observed dark matter density, however, the mass splitting required is much smaller than the value taken in this work~\cite{darkmatter,Ghiglieri:2020ulj}. The remaining mass parameter can be used to perform a scan in the range appropriate for FCC-ee at the Z pole, $M_N=M_1\in[10,\;80]$ GeV. A few benchmark signals were chosen by selecting values for $|U_{\mu1,2}|\in [1\times10^{-6},\; 1\times10^{-4}]$ and then using the ratios in Fig. \ref{fig:points} to consequently obtain the remaining mixing angles. For each mass point and mixing angle combination, 50000 events were generated using the procedure illustrated before. 

\section{Background simulation}
\label{sec:bkg}

\begin{figure}[h]
    \begin{subfigure}[h]{0.5\textwidth}
        \centering
        \includegraphics[width=\textwidth]{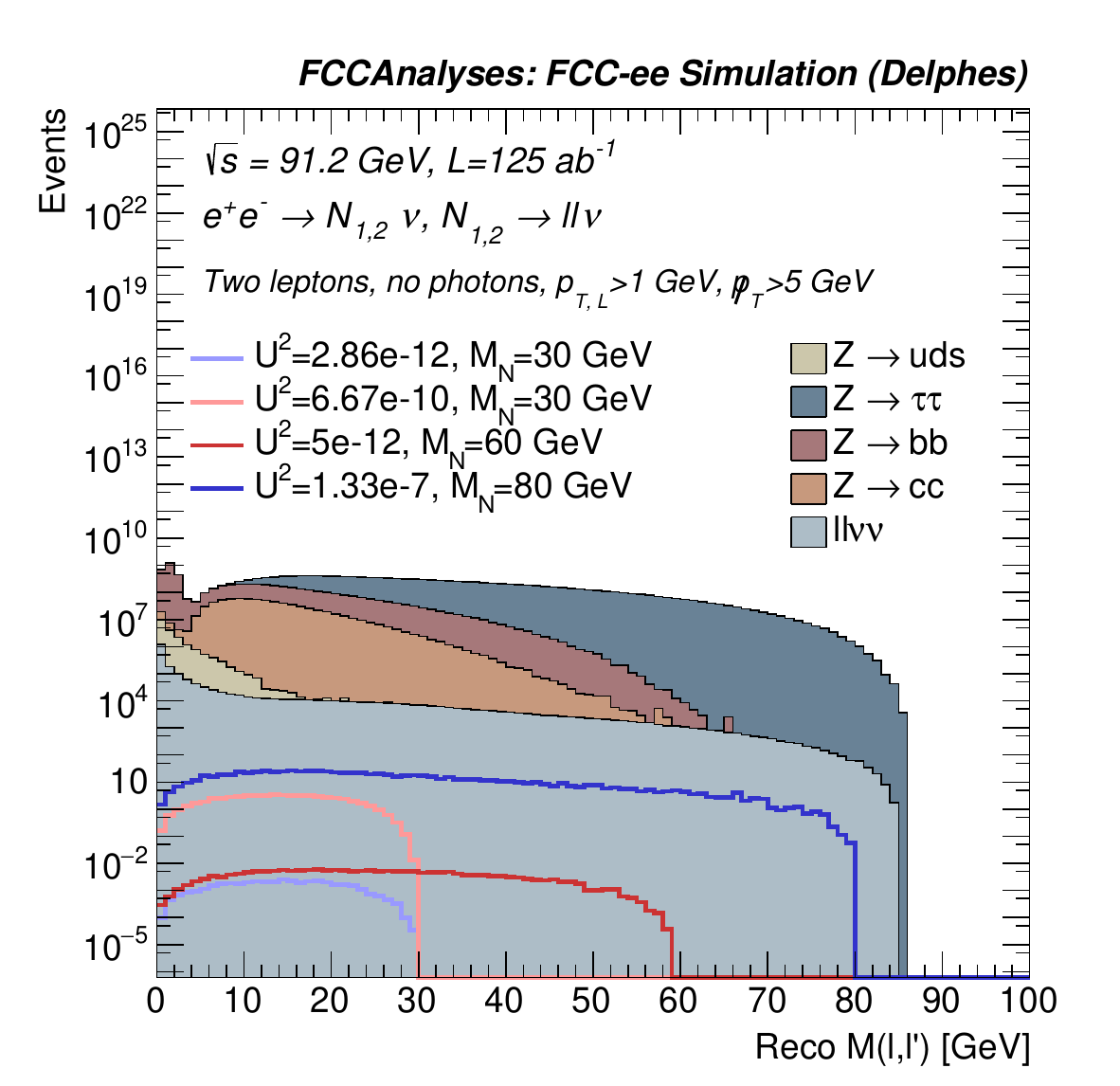}
    \end{subfigure}
    \hfill
    \begin{subfigure}[h]{0.5\textwidth}
        \centering
        \includegraphics[width=\textwidth]{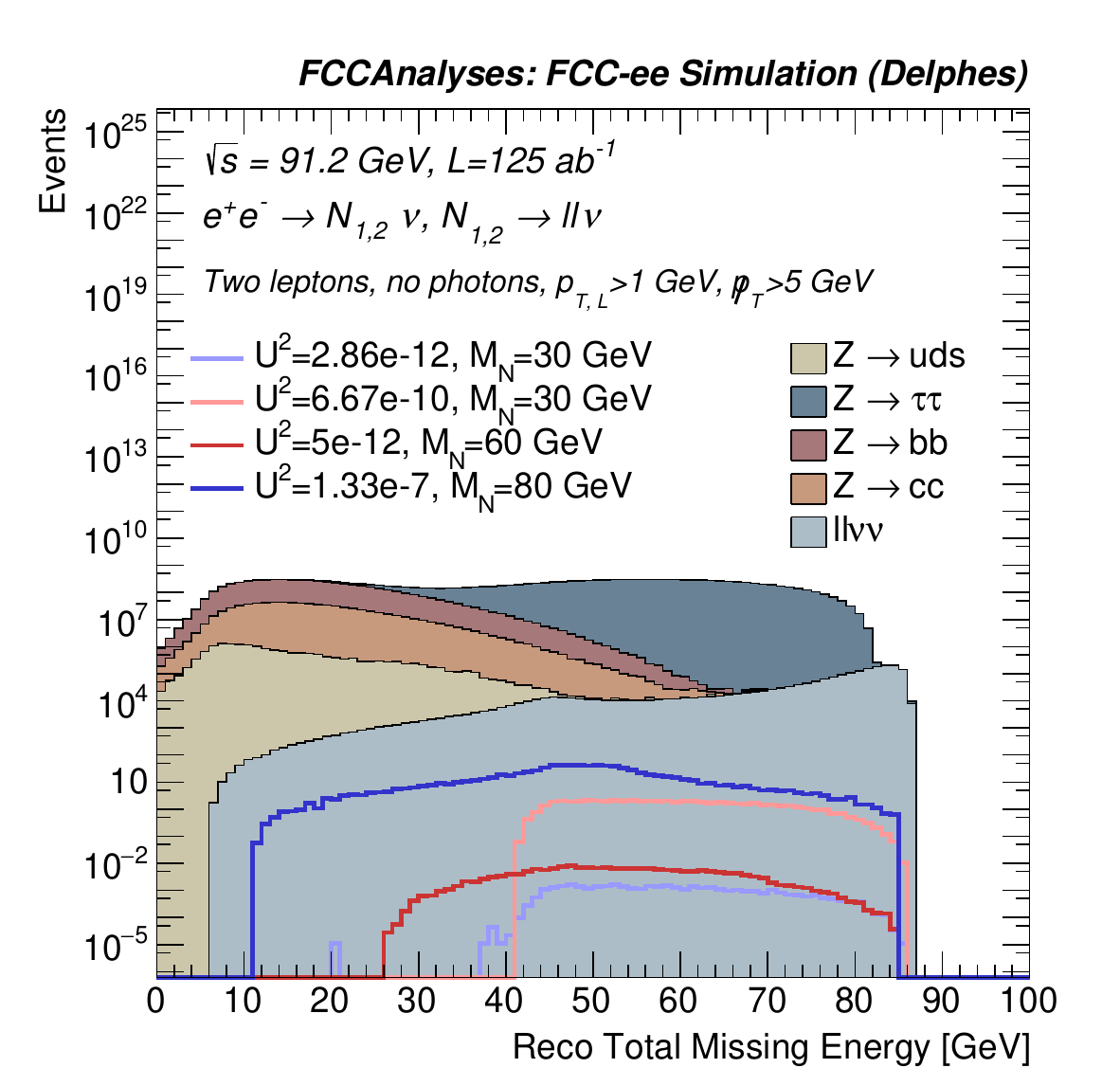}
    \end{subfigure}
    \hfill
    \caption{On the left: invariant mass of the generated di-lepton system. On the right: distribution of the missing energy of the events. The background processes and some benchmark signals are plotted selecting events with two leptons, no photons, and generator cuts.}
    \label{fig:bkgs}
\end{figure}

The main source of background for processes at the Z pole comes from s-channel Z decays, in particular the hadronic decays and $Z\to\tau\tau$. Instead, $Z\to\ell\ell$, where $\ell=e\;\mu$, gives rise to a very small amount of missing energy in the event which makes it irrelevant in this analysis. These processes have been centrally produced within the FCC-ee campaign~\cite{winter23}, using the same simulation setup described for generating the signal samples. Another background source is given by the SM processes producing the same final state as the HNL signal, i.e. $\ell\ell\nu\nu$. This background has been privately generated with Madgraph5\_aMC v3.5.3, using the UFO model \textit{sm-lepton\_masses} which computes amplitudes assuming massive fundamental particles. In particular, the following final states have been generated: $ee\nu\nu$, $\mu\mu\nu\nu$, $\ell\ell'\nu\nu$ where $\ell\neq\ell'$.
For the latter, the diagrams including $e^+e^-\to Z/\gamma* \to \tau \tau$, where one of the tau decays leptonically, were excluded as they are included in the centrally produced Pythia samples. Hadronization and detector response were simulated with Pythia8 and Delphes. To avoid divergences in the cross-section calculation, generator cuts had to be applied for the integration of $\ell\ell\nu\nu$ to proceed correctly in Madgraph: $E>2$ GeV and $p_T>1$ GeV for all charged leptons and $\slashed{p}_T>5$ GeV. These cuts have been propagated to all the signal and background processes analyzed to have a consistent event description. The first cut is aligned with the lepton acceptance of the IDEA detector parameterized in the Delphes card and is therefore implicitly accounted for. 
Fig. \ref{fig:bkgs} shows the distribution of reconstructed level variables for some signal samples and background sources.

\section{Analysis strategy}
\label{sec:analysis}

Two event selections that cover different regions of the parameter space are described in this section: the first one is an inclusive approach, while the second one specifically targets long-lived HNLs.

\subsection{Inclusive analysis event selection}
\label{sec:inclusive}

\begin{figure}[h]
        \centering
        \includegraphics[width=0.5\textwidth]{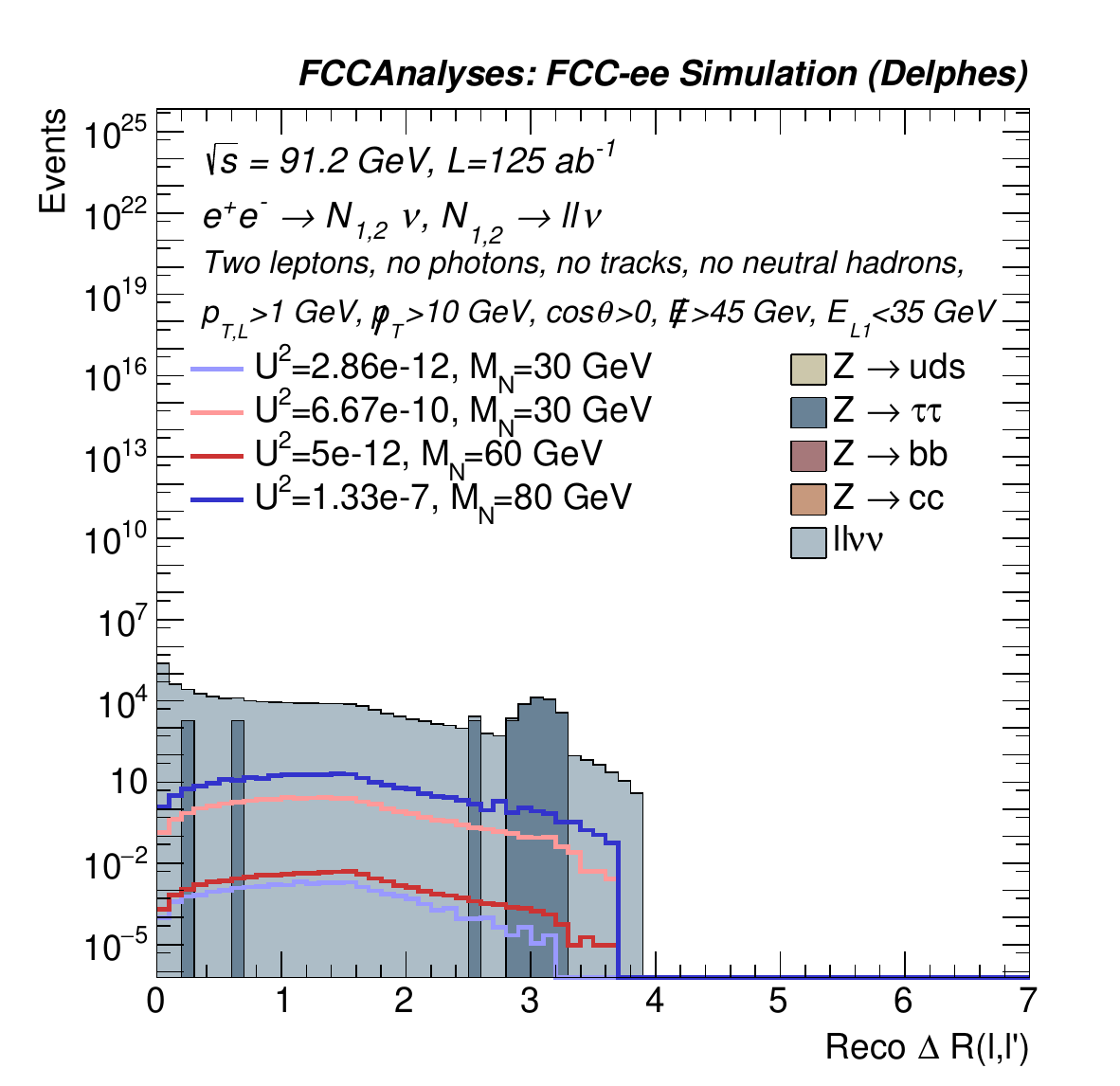}
    \hfill
    \caption{Distribution of the reconstructed angular distance between the two final state leptons. The selected events pass all the selection cuts up to number 7. in Tab. \ref{tab:cuts}. The figure shows four benchmark signal hypotheses and all the backgrounds.}
    \label{fig:bkg_dr}
\end{figure}

The first step of the analysis requires the reconstructed final state to be composed of two opposite-charged leptons (electrons or muons) excluding the presence of photons. 
Then, requiring events with no other reconstructed tracks, besides the two leptons, and no neutral hadrons, the semi-leptonic and hadronic decays of tau in both signal and backgrounds are removed. Subsequently, an optimized selection of cuts on kinematical observables helps improve the signal-to-background ratio. In particular, tailoring the selection on the dilepton invariant mass for different signal hypotheses can effectively achieve this goal since the value is closely associated with the HNL mass, Fig. \ref{fig:bkgs}. 
A summary of the selection criteria is shown in Table~\ref{tab:cuts}. Figure~\ref{fig:bkg_dr} shows the events distributions after the angular distance selection between the two leptons.

\begin{table}[H]
\centering
    \begin{tabular}{lc}
    \hline\hline
    Cut description & Values \\
    \hline\hline
    1. Selection and generator cuts & Two leptons with opposite charges, no photons \\ 
     & $\slashed{p}_{T}>5$ GeV, $p_{T, \ell}>1$ GeV, $E_\ell>2$ GeV \\ 
    2. Reconstructed tracks & No other tracks\\ 
    3. Neutral hadrons & No neutral hadrons with $E>2$ GeV \\ 
    4. Missing transverse momentum & $\slashed{p}_T>10$ GeV \\ 
    5. Cosine between the leptons & $\cos\theta>0$  \\ 
    6. Missing energy & $\slashed{E}>45$ GeV \\ 
    7. Leading lepton energy & $E_{leading\; \ell}<35$ GeV \\ 
    8. Invariant mass & $M(\ell, \ell')<M_{HNL}$ \\ 
    \hline\hline
    \end{tabular}
    \caption{Table of cuts for the inclusive analysis.}
    \label{tab:cuts}
\end{table}

\subsection{Displaced event selection}
\label{sec:displaced}

\begin{figure}[h]
        \centering
        \includegraphics[width=0.5\textwidth]{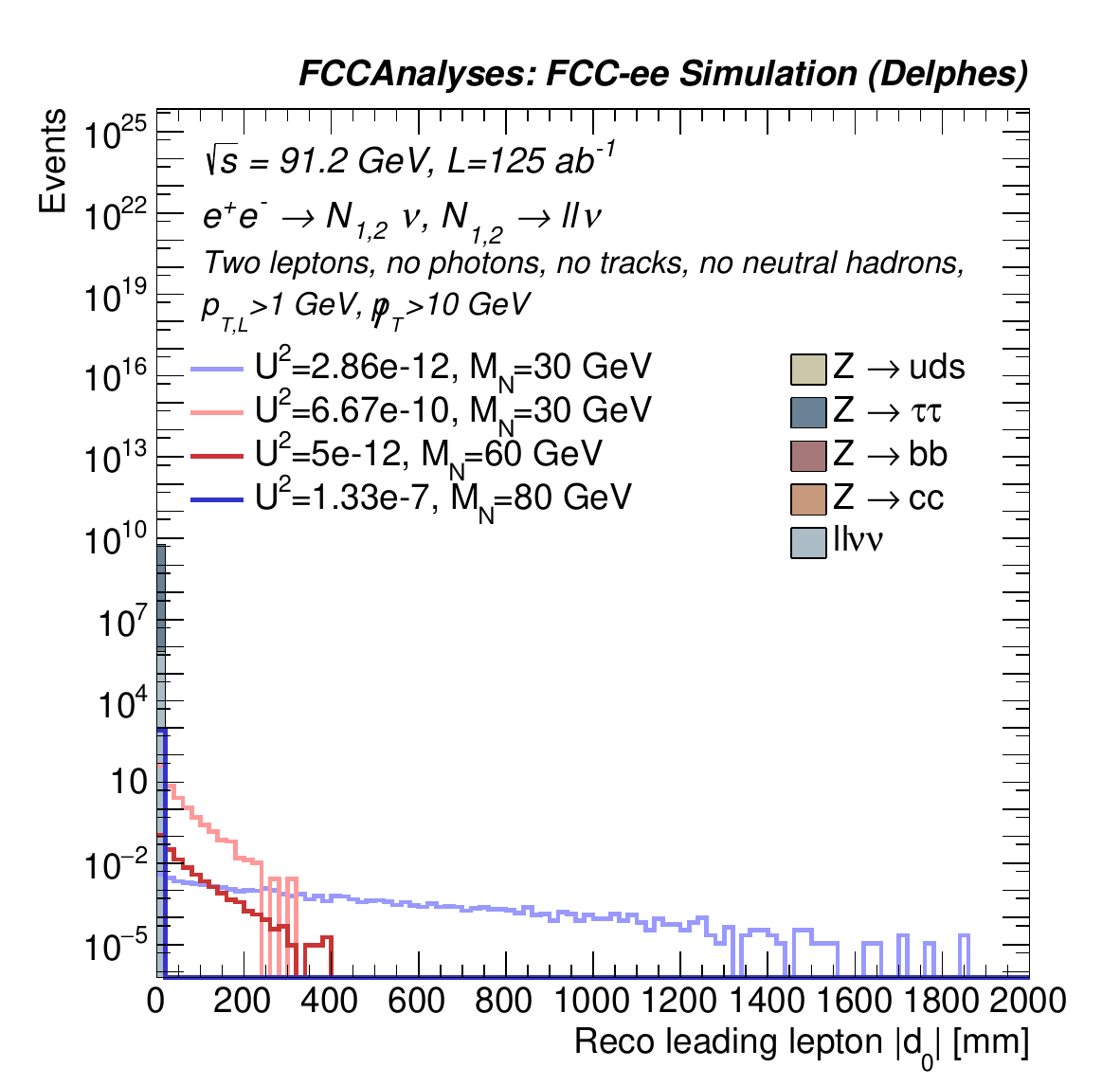}
    \caption{Distribution of the reconstructed impact parameter of the leading lepton in the event. The events selected are those passing the selection criteria up to 4. in Table~\ref{tab:cuts_dv}. The figure shows four benchmark signal hypotheses and all the backgrounds.}
    \label{fig:bkg_d0}
\end{figure}

For the process considered in this work, the reconstruction of the vertex from the two final state leptons closely corresponds to the HNL decay vertex and its properties can be used to differentiate between SM processes that are usually prompt while HNLs decay length depends on the model's parameters~\cite{drewestheory}. 
From Fig. \ref{fig:bkg_d0} it is clear how looking at the distribution of the lepton impact parameter after applying some selection, there are background-free signal events associated with displaced tracks.  
In this case, the selection has been reoptimized to profit from the variable connected to the vertex displacement, achieving a background-free scenario in the simulation. The full list of selection criteria is given in Table~\ref{tab:cuts_dv}.

\begin{table}[H]
\centering
    \begin{tabular}{lc}
      \hline\hline
    Cut description & Values \\
    \hline\hline
    1. Selection and generator cuts  & Two leptons with opposite charges, no photons \\ 
     & $\slashed{p}_{T}>5$ GeV, $p_{T, \ell}>1$ GeV, $E_\ell>2$ GeV \\
    2. Reconstructed tracks & No other tracks\\ 
    3. Neutral hadrons & No neutral hadrons with $E>2$ GeV \\ 
    4. Missing transverse momentum & $\slashed{p}_T>10$ GeV \\ 
    5a. Cosine between the leptons & $\cos\theta>-0.8$  \\ 
    6a. Invariant mass & $M(\ell, \ell')<80$ GeV \\ 
    7a. Vertex $\chi^2$ & $\chi^2<10$ \\
    8a. Lepton transverse impact parameter & $|d_0|>0.64$ mm \\ 
    \hline\hline
    \end{tabular}
    \caption{Table of cuts for the displaced vertex analysis. Cuts reoptimized for this search are indicated by the suffix.}
    \label{tab:cuts_dv}
\end{table}

\section{Results}
\label{sec:results}

This section presents the statistical analysis performed on the two phase spaces introduced previously. First, the significance of detecting HNLs based on signal extraction using discriminating variables, where the method focuses on the angular distance between leptons, is discussed. In the second part, an estimate of the expected number of signal events in a purely displaced analysis is given, highlighting the potential detection of long-lived HNLs in a background-free topology. 

\subsection{Sensitivity of the inclusive analysis} 
\label{subsec:significance} 

\begin{figure}[h]
    \centering
    \includegraphics[width=\textwidth]{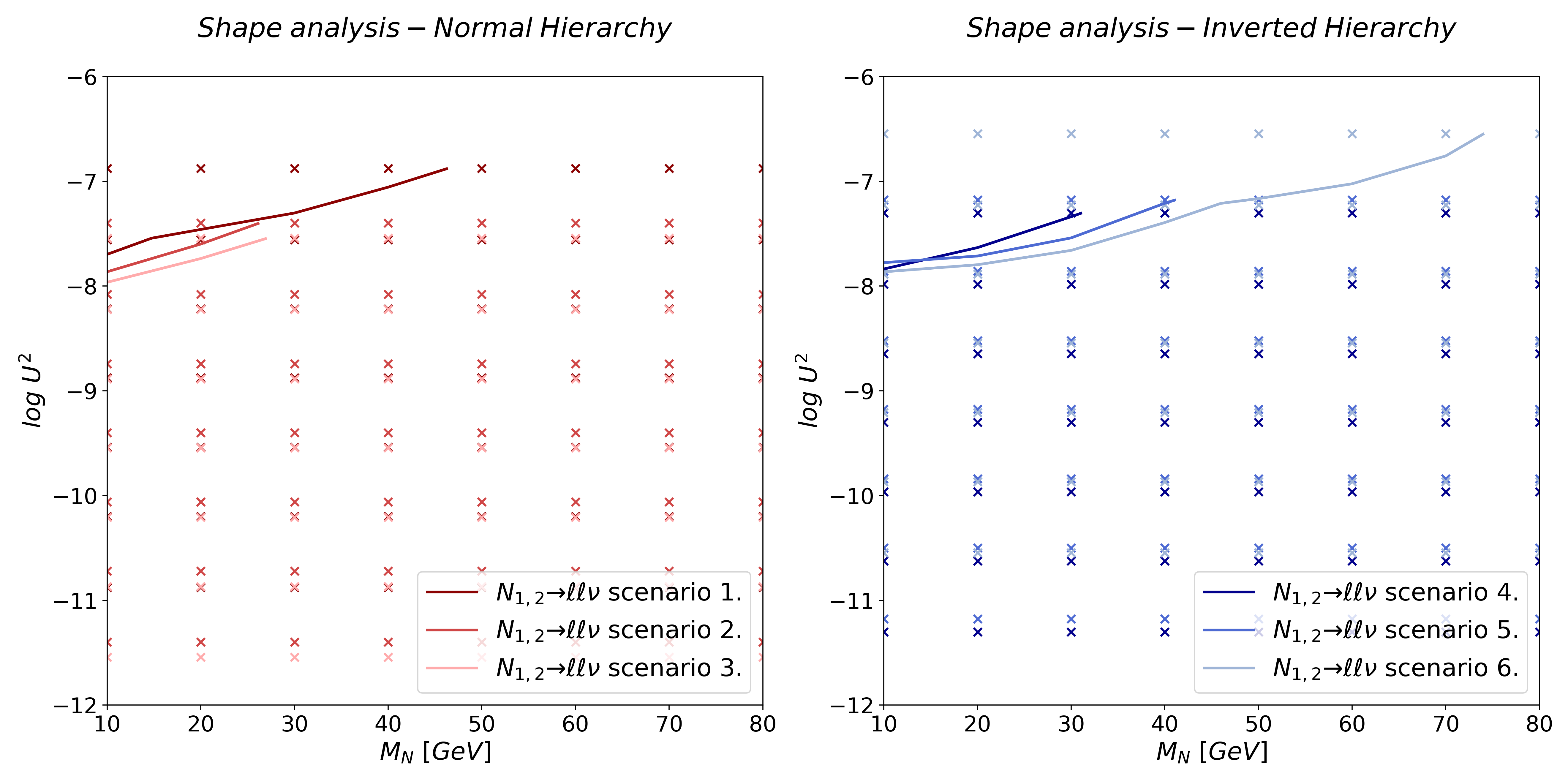}
    \caption{Expected significance of HNLs signals extracted from the angular distance between the two final state leptons, Fig. \ref{fig:bkg_dr}. The plots consider the events passing the cuts in Tab. \ref{tab:cuts}. Contour lines show the regions corresponding to 5 $\sigma$ significance for the different mixing scenarios, illustrated in Fig. \ref{fig:points}, agreeing with normal (left) or inverted (right) neutrino hierarchy.}
    \label{fig:significance}
\end{figure}

The significance for each signal point in the parameter space chosen has been evaluated using the CMS Combine tool~\cite{combine}, analyzing the discriminating variables from the objects reconstructed in the detector simulation in a shape-based analysis. 
The significance is calculated a-priori, independent of any observed datasets. Systematic uncertainties in the Monte Carlo (MC) simulation regarding the predicted cross-section for both signal and background sources are accounted for using a log-normal distribution. Statistical fluctuations in the events are managed by the Barlow-Beeston-lite technique~\cite{barlow-beeston}, which is automatically handled by the CMS Combine software. No additional sources of uncertainty were considered in this calculation. 
The significance has been computed for the events passing the cuts in Tab. \ref{tab:cuts}, using the angular distance between the two final state leptons as a discriminant, Fig. \ref{fig:bkg_dr}. The significance values for different parameter hypotheses have been linearly interpolated between each of the signal samples simulated. The different curves plotted in Fig. \ref{fig:significance} correspond to the points in the parameter space analyzed.
The results demonstrate that for higher mixing angles, the HNL signal can be distinguished from the background with a $5\sigma$ significance, corresponding to prompt HNL signatures. However, this analysis approach does not seem able to further distinguish the various mixing assumptions in case of observation and further refinements would be needed.

\subsection{Displaced signal events}
\label{subsec:displaced_results}

\begin{figure}[h]
    \centering
    \includegraphics[width=\textwidth]{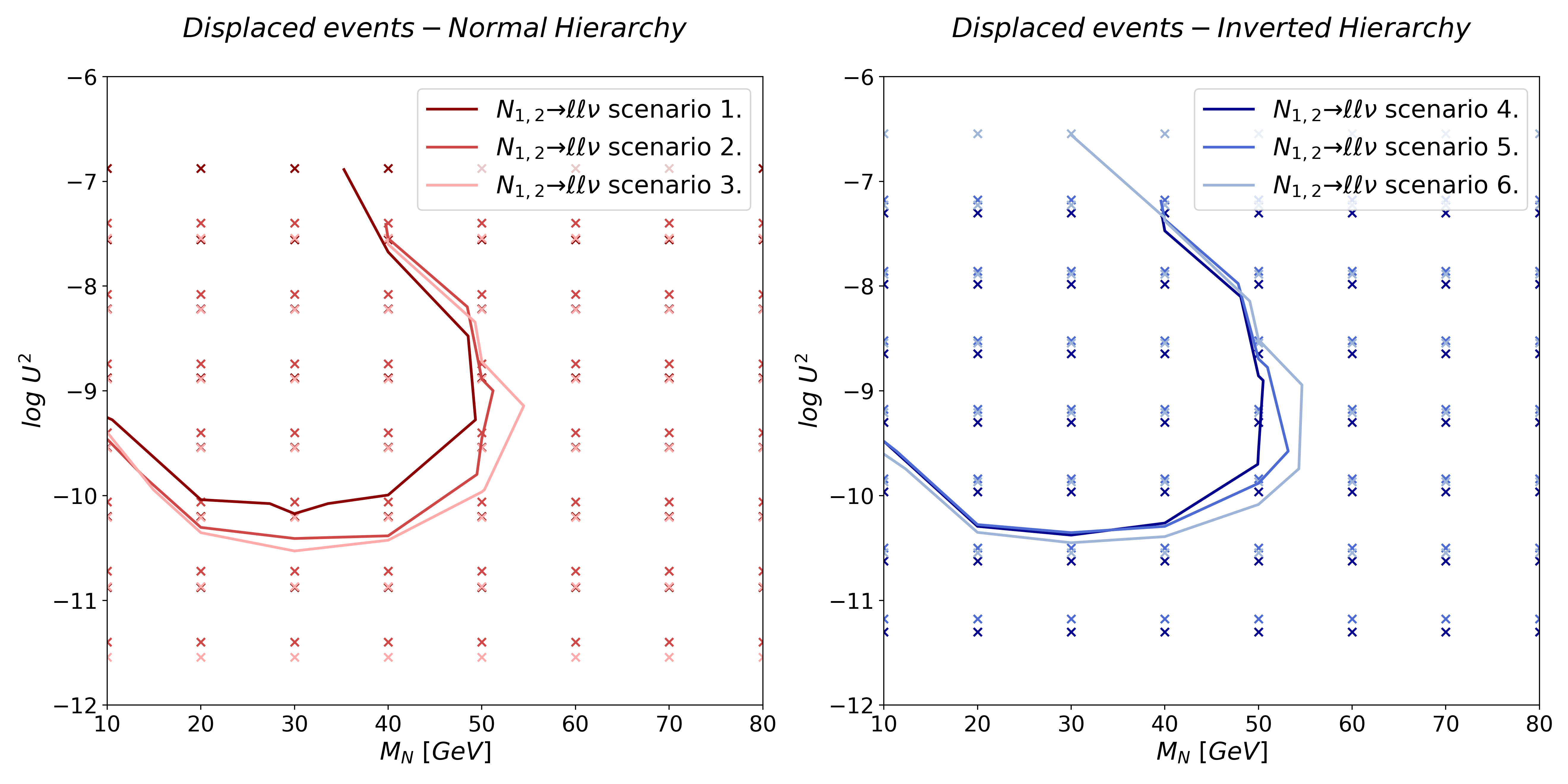}
    \caption{Expected number of long-lived HNLs in a background-free scenario. The plots consider all signal events that pass the cuts on the displaced vertex, Tab. \ref{tab:cuts_dv}. Contour lines show the regions corresponding to at least 4 events for the different mixing scenarios, illustrated in Fig. \ref{fig:points}, agreeing with normal (left) or inverted (right) neutrino hierarchy.}
    \label{fig:n_events}
\end{figure}

In the case of the displaced analysis illustrated in Section \ref{sec:displaced}, the lack of a model for the backgrounds prevents the evaluation of the significance. However, it is possible to evaluate the number of expected signal events, Fig. \ref{fig:n_events}, interpolated linearly between the samples considered in this work to understand the reach of the model. 
For long-lived HNLs, assuming no background events are left after the cuts are applied, a large area of the parameter space could give rise to a sizable amount of signal events. Once again, the lines plotted, choosing the values where more than 4 events would be observed, show that the different mixing assumptions of the HNLs give similar results among them, and further studies would be needed to disentangle them. This analysis needs to be evaluated by taking into consideration the limited modeling of the SM background and the contribution from experimental backgrounds as well.  

\section{Summary}
\label{sec:summary}

Heavy neutral leptons are promising candidates to explain neutrino masses and oscillations. Most searches for HNLs have been conducted by looking at the type I seesaw mechanism with only one generation of HNLs. This simplified benchmark is useful to study the sensitivity of experiments to new physics signals, but it is not theoretically satisfactory since it is incompatible with the observed neutrino masses and mixing patterns. This paper aims to study the detection possibilities of a model containing two generations of HNLs in the type I seesaw mechanism at FCC-ee. 
This choice allows the simultaneous explanation of the order of the observed neutrino masses, the neutrino flavor oscillations, and the generation of baryon asymmetry by leptogenesis in the early Universe. 
The search is based on the data collected during the Z pole run of the FCC-ee, with an expected luminosity of 125 ab$^{-1}$. The analysis is restricted to the 
leptonic decays of HNLs and the backgrounds from relevant SM processes have been included. The sensitivity for HNLs, with parameters consistent with the normal and inverse hierarchy of neutrino masses, has been evaluated by analyzing the angular distance of the two final state leptons. The analysis has been extended to focus on a region expected to be free of SM background, where HNLs are long-lived particles, estimating the sensitivity from the number of signal events that could be observed. The results complement each other and confirm that FCC-ee will have sensitivity to observe HNL signatures in a wide region of the parameter space.

\acknowledgments
This work is supported by the Alexander von Humboldt-Stiftung. This project has received support from the European Union’s Horizon 2020 research and innovation program under grant agreement No. 951754.

\bibliography{biblio}{}
\bibliographystyle{JHEP.bst}

\end{document}